\documentclass[eqsecnum,aps,epsfig]{revtex4}
\usepackage{amsmath,amssymb,amsthm,bm}
\usepackage{psfrag}
\usepackage[dvips]{graphicx}
\usepackage[utf8]{inputenc} 
\usepackage{hyperref}

\begin{document}
\title{Combining the color structures and intersection points of thick center vortices and low-lying Dirac modes}
\author{Seyed Mohsen Hosseini Nejad}
\email{smhosseininejad@ut.ac.ir}
\affiliation{
Faculty of Physics, Semnan University, P.O. Box 35131-19111, Semnan, Iran}

\begin{abstract}

We investigate several examples of Yang-Mills gauge
configurations containing center vortex structures, including
intersection points between vortices and nontrivial color structures
residing on the vortex world-surfaces. Various topological charge
contributions of the color structures and intersection points are studied in these configurations. Low-lying eigenmodes of the
(overlap) Dirac operator in the presence of these vortex backgrounds
are analyzed. The results indicate characteristic properties for spontaneous chiral symmetry
breaking. 
\end{abstract}

\maketitle

\section{INTRODUCTION}\label{Sect0}

Numerical simulations have indicated that the center vortices can account for the phenomena of the color confinement \cite{DelDebbio:1996mh,DelDebbio:1997ke,Kovacs:1998xm,Greensite:2003bk,Vinciarelli:1978kp,Yoneya:1978dt,Cornwall:1979hz,Mack:1978rq,Nielsen:1979xu} and spontaneous chiral symmetry breaking~\cite{deForcrand:1999ms,Alexandrou:1999vx,Engelhardt:2002qs,Hollwieser:2008tq,Reinhardt:2000ck}. 
 Center vortices can contribute to the topological charge through
intersection and writhing points \cite{Reinhardt:2000ck, Reinhardt:2001kf,Engelhardt:2000wc,Bertle:2001xd,Bruckmann:2003yd,Engelhardt:2010ft,Hollwieser:2011uj,Hollwieser:2014mxa,Nejad:2015aia,Hollwieser:2015koa,Nejad:2016fcl}, and
their color structure \cite{Jordan:2007ff,Hollwieser:2010zz,Hollwieser:2012kb,Schweigler:2012ae,Nejad:2015aia,Nejad:2016fcl}. We have studied colorful plane vortices in Refs. \cite{Nejad:2015aia,Nejad:2016fcl}. According to the Atiyah-Singer index theorem \cite{Atiyah:1971rm,Schwarz:1977az,Brown:1977bj,Adams:2000rn}, the zero modes of the Dirac operator are related to the topological charges. Any source of topological charge
can attract (would-be) zero modes which contribute via interactions in Monte-Carlo simulations to a finite
density of near-zero modes. By the Banks-Casher relation \cite{Banks:1979yr}, the finite
density of near-zero modes leads to a non-zero chiral condensate and therefore chiral symmetry breaking. For studying the contributions of different types of topological charge sources of center vortices
and analyzing low-lying modes of Dirac operator for vortex configurations with these topological charge sources, it is necessary to consider special vortex configurations. The color structures and intersection points of center vortices producing the topological charge may appear on the vacuum, simultaneously.

In this article, we study some plane vortex configurations in SU($2$) lattice gauge theory which are combinations of the color structures and intersection points. These configurations give the nice opportunity to study the properties of zero modes and near-zero modes. On periodic lattices, plane vortices appear in parallel or antiparallel pairs. By combining two perpendicular vortex pairs and making one of the plane vortices colorful, we get some configurations where the intersection points and color structures are apart. Various topological charge
contributions of the color structures and intersection points are studied in these configurations. We calculate eigenvalues and low-lying eigenmodes of the overlap Dirac operator in the background of the considered vortex configurations and the chiral properties of the low-lying eigenmodes are analyzed.

In Sect.~\ref{Sect1} we describe the planar vortex configurations. Combinations of intersections and colorful regions and topological charges of these vortex configurations are investigated in Sect.~\ref{Sect2}. Then, in Sect.~\ref{Sect3}, we study eigenvalues and eigenmodes of the overlap Dirac operator for the vortex configurations and analyze the influence of combinations of intersection points and colorful regions on low-lying modes of the Dirac operator. In the last step, Sect.~\ref{Sect4}, we summarize the main points of our study. 

\section{thick planar vortex configurations}\label{Sect1}

In SU($2$) lattice gauge theory, we investigate thick plane vortices \cite{Hollwieser:2011uj,Jordan:2007ff} which are parallel to two of the coordinate axes and occur in pairs of parallel sheets because of using periodic  boundary conditions for the gauge fields. We use vortex plane pairs which extend in $zt$- and $xy$-planes. The lattice links of plane vortices as an unicolor vortex field vary in a $U(1)$ subgroup of SU($2$) and usually are defined by the Pauli matrix $\sigma_3$, {\it i.e.} $U_\mu=\exp\{\mathrm i\alpha \sigma_3\}$. For $xy$-vortices ($zt$-vortices), $t$-links ($y$-links) are nontrivial in one $t$-slice $t_\perp$ ($y$-slice $y_\perp$). The orientation of the vortex flux is determined by the gradient of the angle $\alpha$ which is chosen as a linear function of the coordinate perpendicular to the plane vortices. For $xy$-vortices, the appropriate profile functions for angle $\alpha$ are given as the following:
\begin{equation}
 \alpha_1(z) = \begin{cases}     2\pi \\ \pi\left[ 2-\frac{z-(z_1-d)}{2d}\right] \\ 
                           \pi \\ \pi\left[ 1-\frac{z-(z_2-d)}{2d}\right] \\ 
                          0 \end{cases} \ldots
  \alpha_2(z) = \begin{cases}     0 & 0 < z \leq z_1-d \\ 
                \frac{-\pi}{2d}[z-(z_1-d)] & z_1-d < z \leq z_1+d \\ 
                -\pi & z_1+d < z \leq z_2-d \\ 
                -\pi\left[ 1-\frac{z-(z_2-d)}{2d}\right] & z_2-d < z \leq z_2+d \\ 
                0 & z_2+d < z \leq N_z \end{cases}. \label{eq:phi-pl0}
\end{equation}

 The parallel sheets of $xy$-vortices have thickness of $2d$ centered around $z_1$ and $z_2$. Vortex pairs with
the same (opposite) vortex flux orientation corresponding to the angle $\alpha_1$ ($\alpha_2$) are called parallel (anti-parallel) vortices. It should be noted that the unicolor vortices defined with
Eq.~(\ref{eq:phi-pl0}) are not thickened in both transverse dimensions, an $xy$-vortex is thick in $z$-direction, but still thin in $t$-direction. Thus, these
vortices resemble bands rather than tubes. The vortex profiles $\alpha(x)$ similar to Eq.~(\ref{eq:phi-pl0}) are used for $zt$-vortices centered around $x_1$ and $x_2$. An $zt$-vortex is thick in $x$-direction, but still thin in $y$-direction.

The topological charge of the configurations on the continuum is defined as:
\begin{equation}\label{eq:qlat}
  Q = \int d^4x \, q(x)= -\frac{1}{16\pi^2}\int d^4x \, \mbox{tr}[\tilde{\cal F}_{\mu\nu} \cal F_{\mu\nu} ]
\end{equation}
where $q(x)$ denotes topological charge density, $\tilde{\cal F}_{\mu\nu}=\frac{1}{2} \epsilon_{\mu\nu\alpha\beta}{\cal F}_{\alpha\beta}$, ${\cal F}_{\mu\nu}=-ig F_{\mu\nu}$, $F_{\mu\nu}=F_{\mu\nu}^a\frac{\sigma^a}{2}$, and $F_{\mu\nu}^a = \partial_\mu A_\nu^a - \partial_\nu A_\mu^a + g f_{abc} A_\mu^b A_\nu^c$. The plaquette definition discretises the continuum expression \cite{DiVecchia:1981aev, DiVecchia:1981hh, hollwieser:2009wka, Jordan:2005}. The trace evaluates to 
\begin{equation}
 \mbox{tr}[\tilde{\cal F}_{\mu\nu} {\cal F}_{\mu\nu} ] = (-2g^2)[F_{12}^a F_{34}^a - F_{13}^a F_{24}^a + F_{14}^a F_{23}^a] \label{eq:ftr}
\end{equation}
The field tensor $F_{\mu\nu}$ is obtained from the plaquette variables $U_{\mu\nu}= U_{\mu}(x)U_{\nu}(x +\mu)U^\dagger_{\mu}(x+\nu)U^\dagger_{\nu}(x)$ as
\begin{equation}
  U_{\mu\nu}(x) \approx \exp\left[i a^2 g F_{\mu\nu}^b (x) \frac{\sigma^b}{2} \right].
\end{equation}
If a plaquette is written as
\begin{equation}
  U_{\mu\nu}(x) = u_{\mu\nu}^0(x) I + i u_{\mu\nu}^b(x) \sigma^b,
\end{equation}
 one can obtain
\begin{equation}
  F_{\mu\nu}^b =  \frac{2}{a^2 g}u_{\mu\nu}^b(x). \label{eq:fsin}
\end{equation}
To define a field tensor $F_{\mu\nu}$ for a single lattice point, we average over the four adjacent plaquettes in forward and backward direction. This means we make in the above expressions the substitution
\begin{equation}
  U_{\mu\nu}(x) \to \tilde U_{\mu\nu}(x) = \frac{1}{4} \sum U_{\pm\mu,\pm\nu}(x).
\end{equation}
Combining Eqs.~(\ref{eq:qlat}), (\ref{eq:ftr}), and (\ref{eq:fsin}), the lattice topological charge density is \cite{Jordan:2005}
\begin{equation}
  q(x) = \frac{1}{2\pi^2 a^4} [\tilde u_{12}^a \tilde u_{34}^a - \tilde u_{13}^a \tilde u_{24}^a + \tilde u_{14}^a \tilde u_{23}^a],\label{eq:qu}
\end{equation}
where $\tilde u_{\mu\nu}$ are the components of $\tilde U_{\mu\nu}$. Therefore, the lattice topological charge is obtained as
\begin{equation}
  Q = a^4 \sum_x q(x).
\end{equation}

For the plane vortices as the unicolor vortex fields, one gets vanishing gluonic topological charge. The color structure and intersection points of the plane vortices can contribute to the topological charge. 

Each intersection between $xy$- and $zt$-vortices carries a topological charge with modulus $|Q| = 1/2$, whose sign depends on the
relative orientation of the vortex fluxes \cite{Hollwieser:2011uj}. 

A colorful region of the $xy$-vortex with radius $R$ around $(x_0,y_0)$ at $z_1$ and $t_\perp$ is defined by the time-like links $\exp\{\mathrm i\alpha\vec n\vec\sigma\}$ distributed over the full SU($2$) gauge group \cite{Nejad:2015aia}. The color direction $\vec n$ is 
\begin{equation}
\vec{n}=\sin \left( \theta\left(\rho\right)\right)\cos \left( \phi\right)\hat{i}+\sin \left( \theta\left(\rho\right)\right)\sin \left( \phi\right)\hat{j}+\cos \left( \theta\left(\rho\right)\right)\hat{k},
\end{equation}
where
\begin{equation}\begin{aligned}\label{DefColVort}
&\rho=\sqrt{(x-x_0)^2+(y-y_0)^2},\quad
\theta(\rho)=\pi(1-\frac{\rho}{R})H(R-\rho)\;\in\,[0,\pi],\quad
\phi=\arctan\frac{y-y_0}{x-x_0}\;\in\,[0,2\pi).
\end{aligned}\end{equation}
$H$ is the Heaviside step function. For the colorful configuration as defined up to this point, one obtains vanishing gluonic topological charge \cite{Nejad:2015aia}, but this can be seen to represent a lattice artifact. By a gauge transformation, one can show that the colorful configuration represents a fast transition between two vacua with different winding number in temporal direction (sharp vortices) \cite{Schweigler:2012ae,Nejad:2015aia}. 
After smoothing the colorful vortices by growing the temporal extent $\Delta t$ of the configuration, the topological charge of the configuration becomes non zero. The topological charge of the smoothed configuration obtains modulus $|Q| = 1$, whose sign depends on the
profile of the colorful vortex sheet. Therefore, for the colorful vortices, it turns out to
be quite crucial to thicken in both transverse dimensions and vanishing of the topological charge of the sharp vortices ($\Delta t=1$) is a lattice artifact due to the singularity of the vortex in temporal direction. 

In the following, we study combinations of the color structure and intersection points of center vortices.  

\section{Combinations of colorful regions and intersection points}\label{Sect2}

The color structures and intersection points of center vortices may appear on the vacuum, simultaneously. By combining these sources of topological charge, we investigate the influence of these combinations on low-lying modes of the Dirac operator. According to the topological charge definition
\begin{gather}\label{eq:qlatq}
  Q=-\frac{1}{32\pi^2}\int d^4x\,\epsilon_{\mu\nu\alpha\beta}\,
\mbox{tr}[{\cal F}_{\alpha\beta} {\cal F}_{\mu\nu} ]
=\frac{1}{4\pi^2}\int d^4x\,\vec E^a\cdot\vec B^a,
\end{gather}
when a configuration has electric and magnetic fields, it can contribute to the topological charge. The index $a$ is related to the three directions $\sigma_a$ of the SU($2$) color
algebra. The unicolor regions of plane vortices, as explained in previous section, are defined by the Pauli matrix $\sigma_3$. The unicolor region of xy-vortices has only nontrivial zt-plaquettes and therefore an electric field $E_{z}^3$, while the one of zt-vortices bears
nontrivial xy-plaquettes and therefore a magnetic field $B_{z}^3$. The topological charge of an intersection point between xy- and zt-vortex sheets is then proportional to $E_{z}^3B_{z}^3$. The intersection points with the same (opposite) orientation of xy- and zt-vortex fluxes are parallel (anti-parallel) crossings and their topological charge contribution is $Q = 1/2$ ($Q = -1/2$).
 
For the colorful regions, three directions of $\sigma_a$ contribute to the topological charge while for the intersection points one direction of $\sigma_a$ contributes to the topological charge. 

Now, we study some configurations that the colorful regions and intersection points are combined. In the configurations, $xy$-vortices which have the color structures are thickened in both transverse dimensions but $zt$-vortices are thickened in $x$-direction, but still thin in $y$-direction.

\begin{figure}[h!]
\centering
\includegraphics[width=0.45\columnwidth]{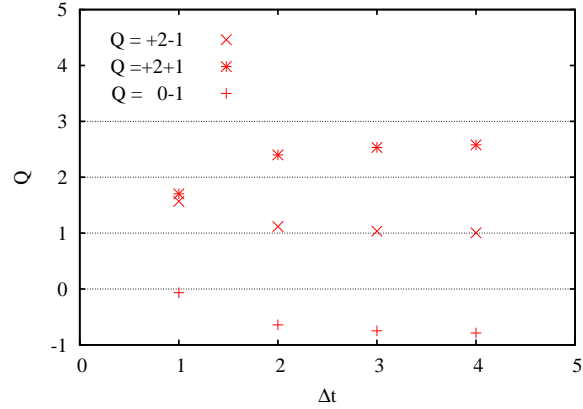}
\caption{ The total topological charge of some vortex configurations corresponding to Fig.~\ref{fig:2} which are combinations of colorful region and intersection points. In the figure, topological charge $Q$ of any configuration is $Q_{int}+Q_{col}$, the sum of intersection and colorful contributions. For the configurations as the fast transition in temporal direction ($\Delta t=1$), the topological charge contribution of colorful region is not appeared due to the singularity of the configuration in temporal direction. After growing temporal extent of the configurations, the topological charge contribution of colorful region is added.}
\label{fig:1}
\end{figure}

\begin{figure}[h!]
\centering
a)\includegraphics[width=0.3\columnwidth]{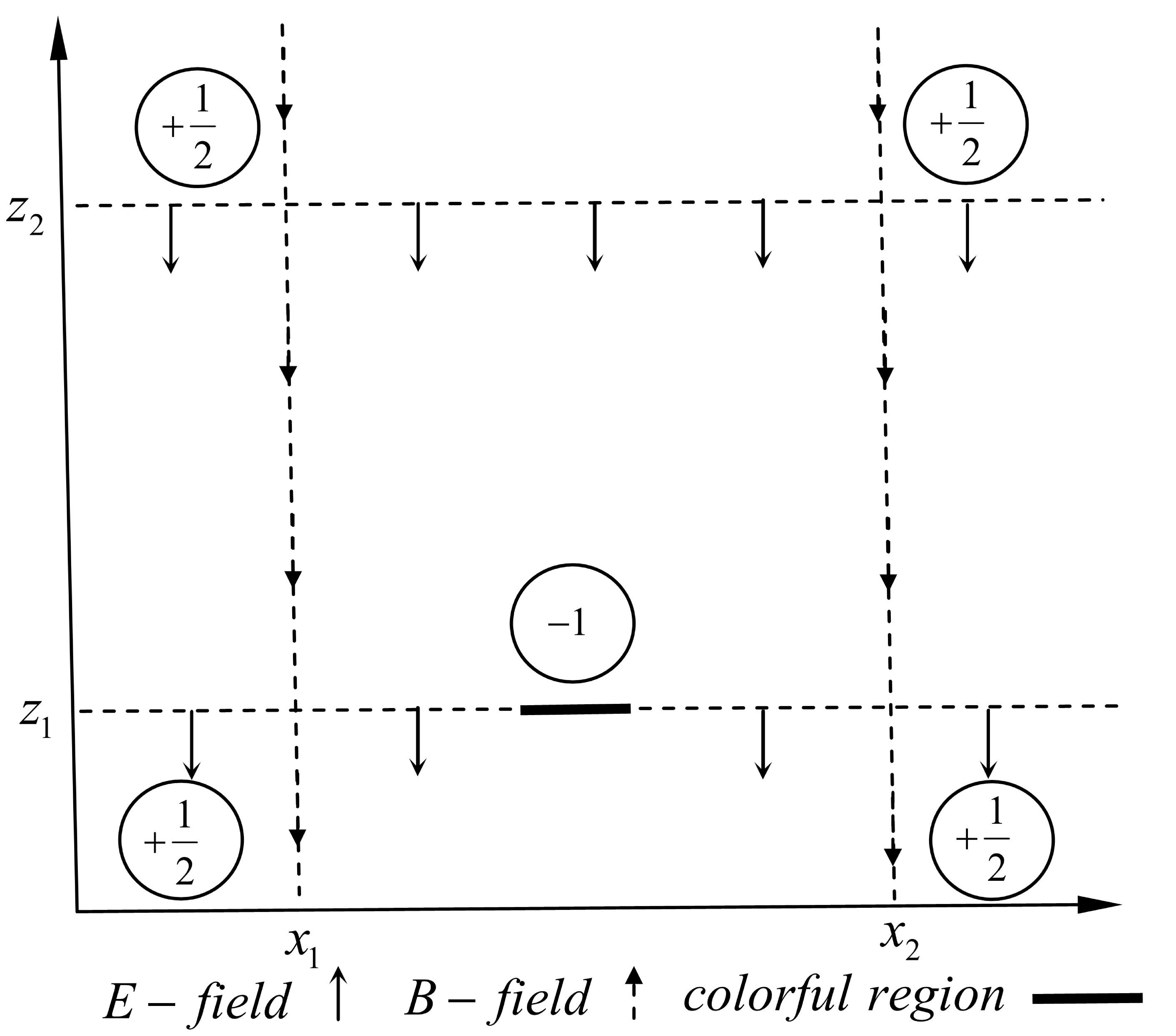}
b)\includegraphics[width=0.3\columnwidth]{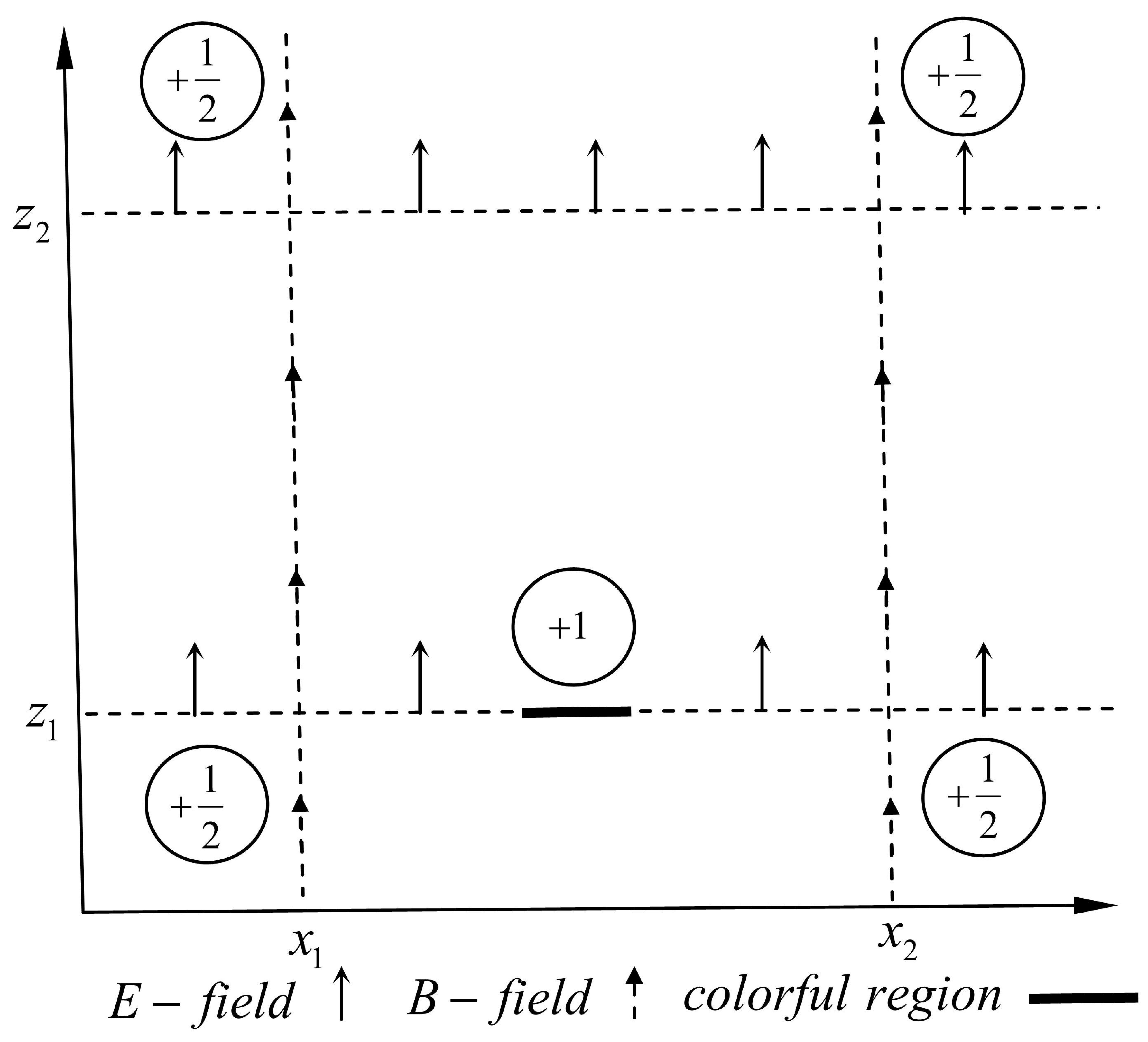}
c)\includegraphics[width=0.3\columnwidth]{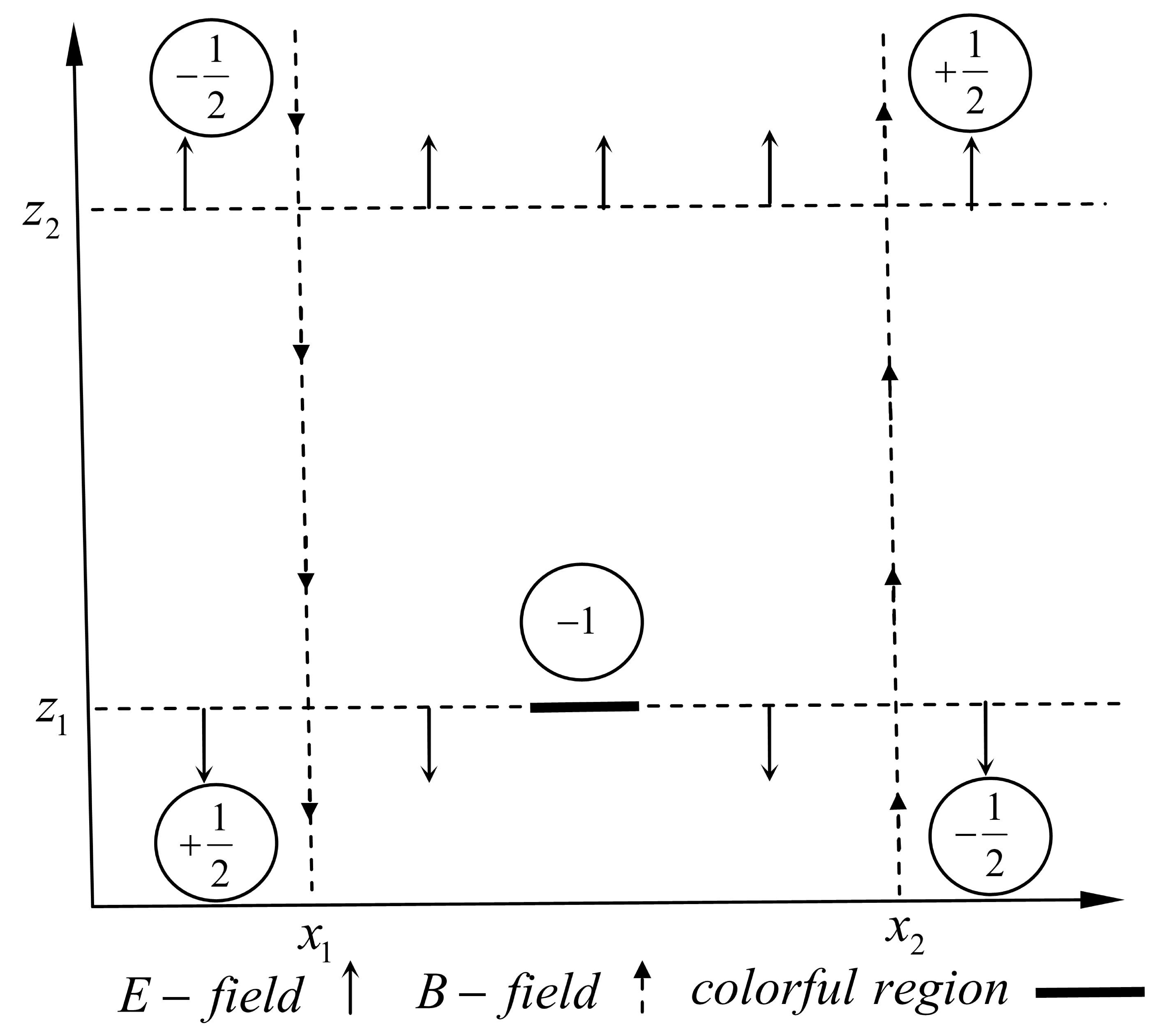}
\caption{ The geometry, field strength and the topological charge contributions of the intersection points and colorful region in the intersection plane. The bold black line indicates that the unicolor vortex is substituted in this region by a colorful region. a) $Q=1$ configuration where two parallel vortex pairs are intersected. We use $\alpha_1$ as the angle profile of the vortex pairs. The total topological charge contribution of intersection points is $Q=2$ and the one of colorful region is $Q=-1$. Therefore various contributions sum up to a total topological charge $Q=1$.  b) $Q=3$ configuration which is the same as $Q=1$ configuration but the angle profile of the vortex pairs is $-\alpha_1$. The total topological charge contribution of intersection points is $Q=2$ and the one of colorful region is $Q=1$. Therefore, various contributions sum up to a total topological charge $Q=3$. c) $Q=-1$ configuration where two anti-parallel vortex pairs are intersected. We use $\alpha_2$ as the angle profile of the vortex pairs. The total topological charge contribution of intersection points is Q=0 and the one of colorful region is $Q=-1$. Therefore various contributions sum up to a total topological charge $Q=-1$.}
\label{fig:2}
\end{figure}

\begin{figure}[h!] 
\centering
a)\includegraphics[width=0.3\columnwidth]{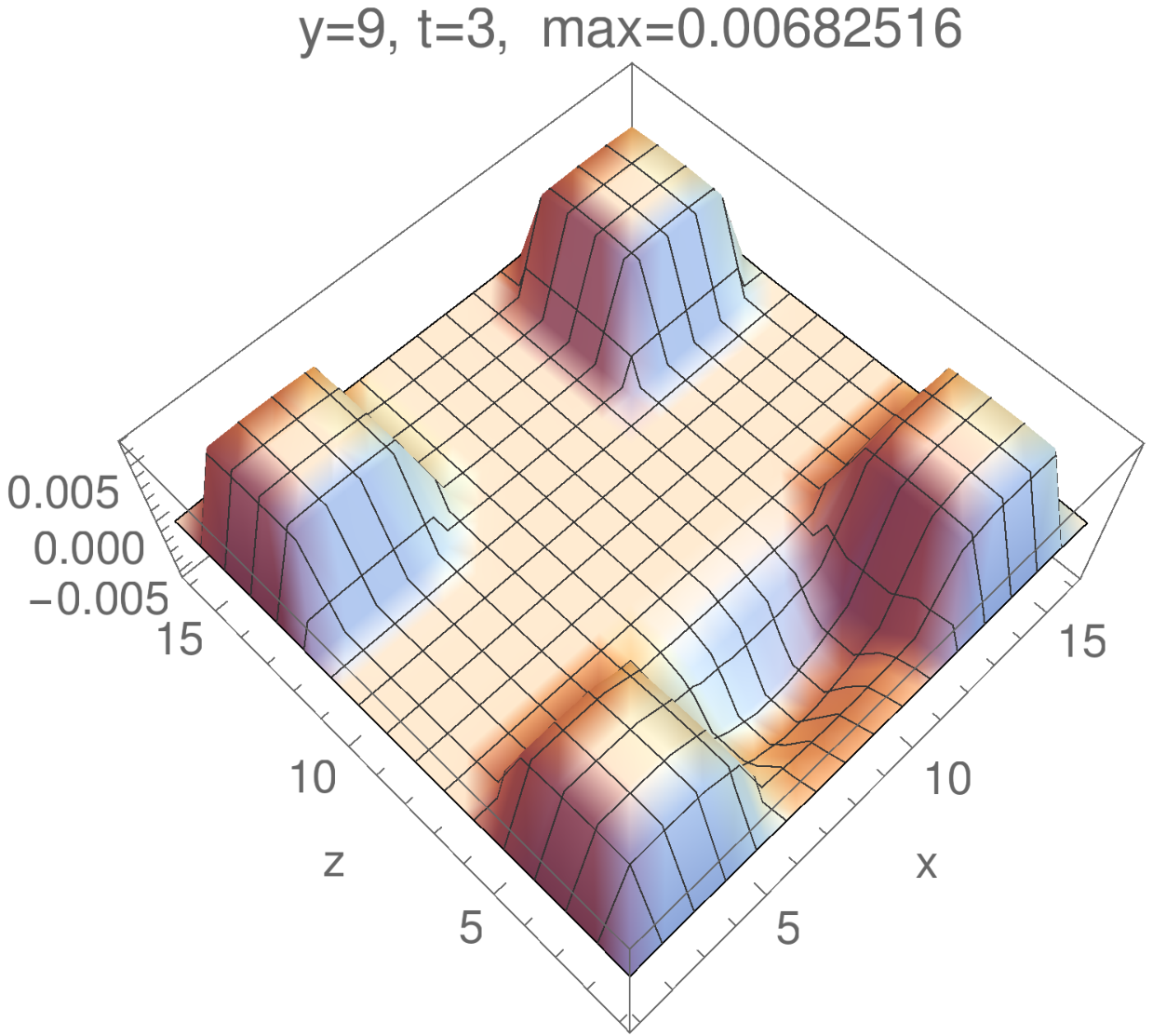}
b)\includegraphics[width=0.3\columnwidth]{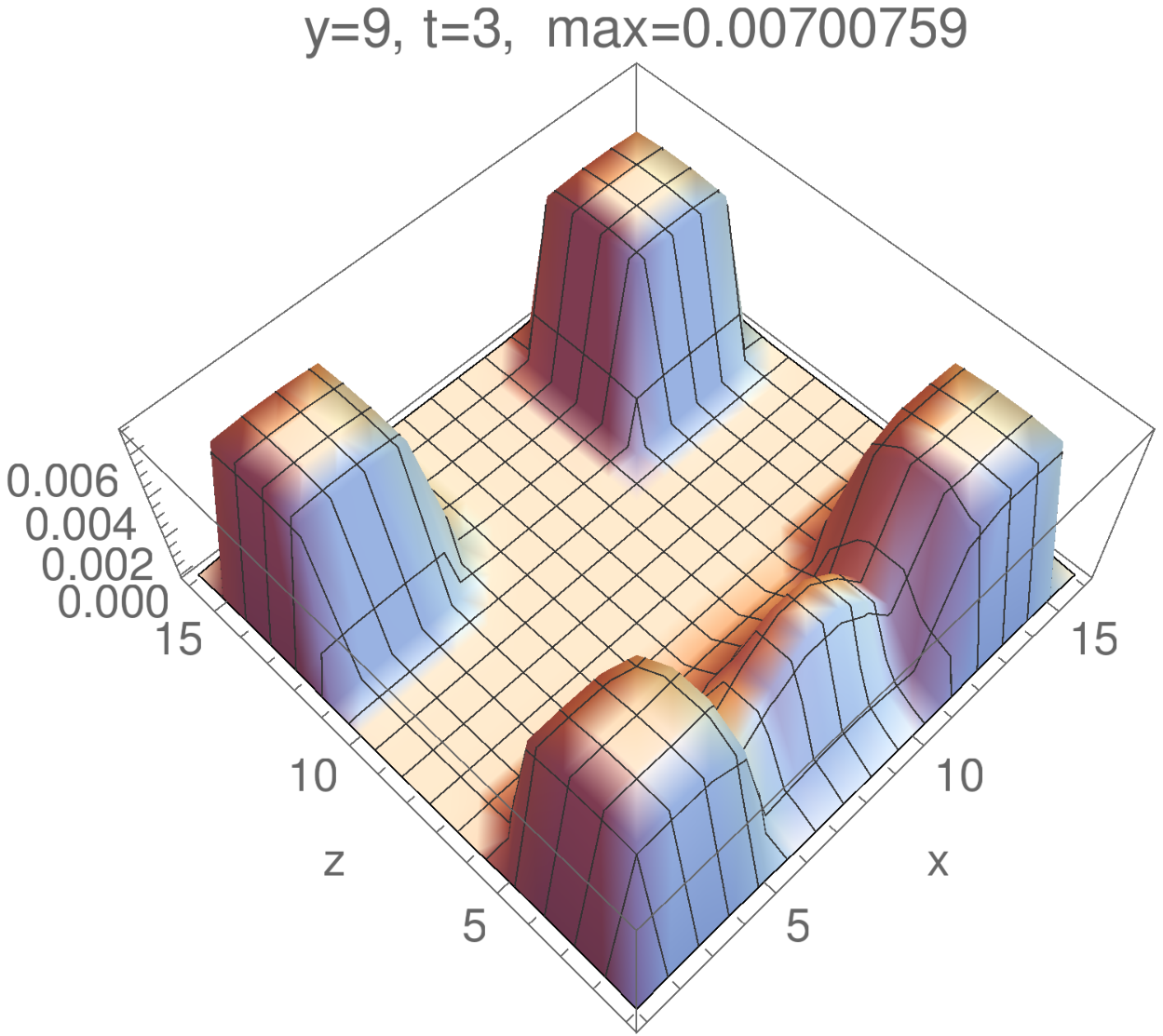}
c)\includegraphics[width=0.3\columnwidth]{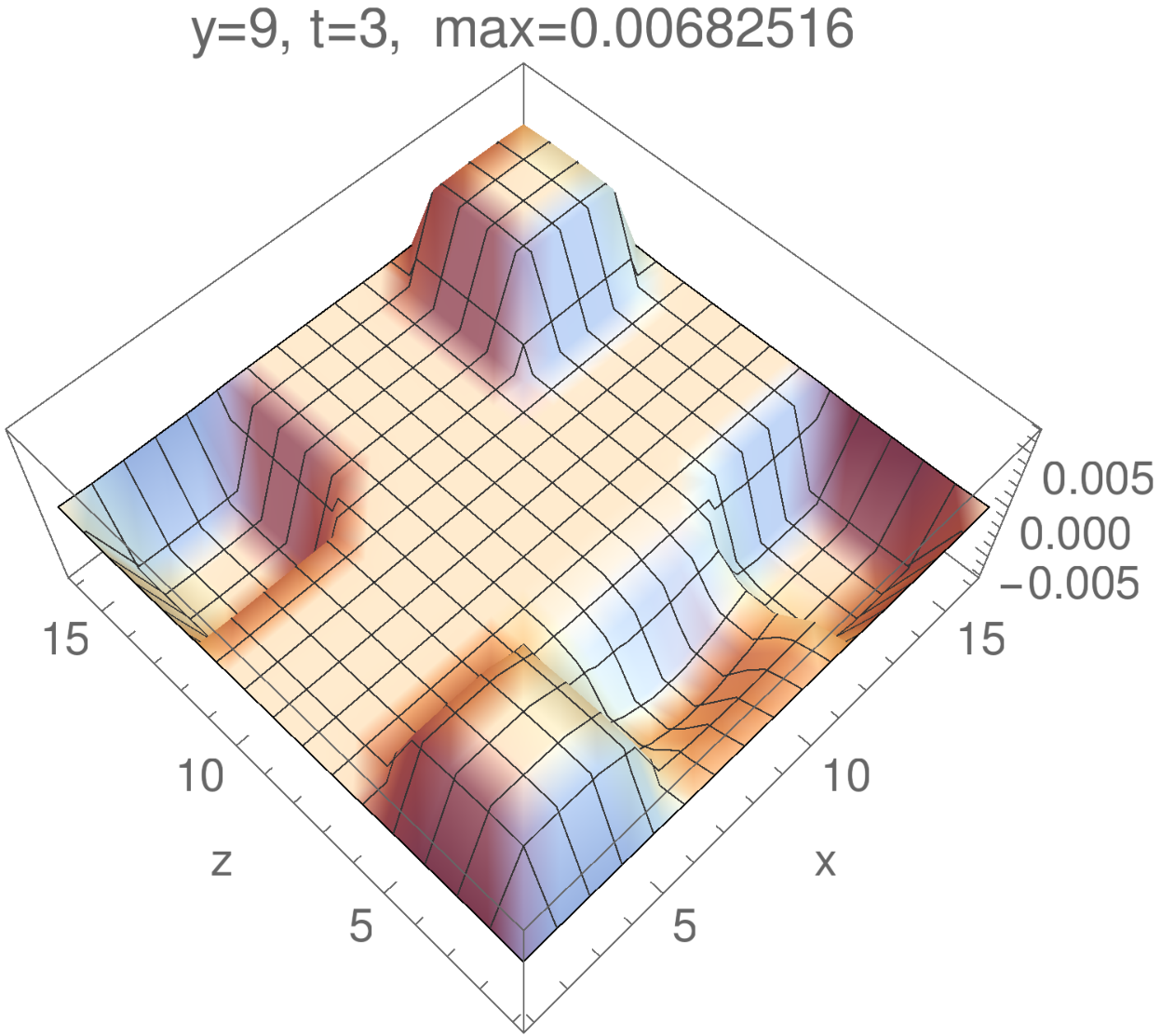}
\caption{The topological charge densities in the $xz$-plane for the configurations displayed in Fig.~\ref{fig:2} which are two intersecting $xy$- and $zt$-plane vortex pairs with vortex centers ($z_1=4,z_2=14$) and ($x_1=4,x_2=14$) at $(t_\perp=3, y_\perp=9)$ with thickness $d = 2$ and $\Delta t = R = 4$ on a $(16)^4 $-lattice. The color structure of the $xy$-vortices is located in the first vortex sheet ($z_1=4$) and the vacuum to vacuum transition occurs between $t = 1$ and $t = 5$. The center of the colorful region is located at $x_0=y_0=9$ in the xy plane. }
\label{fig:3}
\end{figure}

For the first step, we intersect a parallel $xy$-vortex pair at ($z_1=4,z_2=14$) with a parallel $zt$-vortex pair at ($x_1=4,x_2=14$) with thickness $d=2$ at ($t_\perp=3$, $y_\perp=9$) on a $16^4$-lattice. We use $\alpha_1$, given in Eq.~(\ref{eq:phi-pl0}), as the angle profile of the vortex pairs. The colorful region with radius $R = 4$ is located in the first vortex sheet of the $xy$-plane vortices ($z_1=4$) with the center at $x_0= y_0 = 9$ in the xy plane. The intersection points and the colorful region are apart. However there is a quite small overlap between these regions which has only a minor effect. Note that the radius of the colorful region and the thickness of vortex sheets should be appropriate for obtaining the Dirac modes. In Fig.~\ref{fig:1}, the topological charge of the configuration is plotted for various values of $\Delta t$. In the plot, the $Q$ symbol is composed of $Q_{int}+Q_{col}$, the sum of topological charges of intersection and colorful contributions. As shown, for the sharp configuration where $\Delta t=1$, the total charge contribution $Q=2$ of four intersection points appears. The topological charge after smoothing the configuration, where the charge contribution of color structure is added, approaches from $Q=2$ to $Q=2-1=1$. The schematical diagram for the intersection plane of this configuration is shown in Fig.~\ref{fig:2}a. Each intersection point and colorful region carry topological charge $Q=+1/2$ and $Q=-1$, respectively. The minimum value of the action $S$ as a function of $\Delta t$ corresponding to the colorful region is 1.68~$S_\textrm{inst}$ around $\Delta t=R$, where the instanton action $S_\mathrm{inst}=8\pi^2/g^2$ \cite{Nejad:2015aia}. Therefore, we consider $\Delta t=R$, with our choice of $R=4$, cf. above; the transition thus occurs between $t = 1$ and $t = 5$. In Fig.~\ref{fig:3}a, the topological charge
density of the $Q=1$ configuration at $(t_\perp=3, y_\perp=9)$ is plotted in the xz-plane (the intersection plane), where we can identify the positive and negative contributions indicated in Fig.~\ref{fig:2}a.

For the next step, we consider a configuration that is the same as the first configuration but the angle profile of the vortex pairs is $-\alpha_1$. The parameters for this vortex configuration are the same
as those of the $Q=1$ configuration. As shown in Fig.~\ref{fig:1}, for the sharp configuration, the total charge contribution $Q=2$ of four intersection points appears. The topological charge after smoothing the configuration approaches from $Q=2$ to $Q=2+1=3$. The schematical diagram for the intersection plane of this configuration is shown in Fig.~\ref{fig:2}b. Each intersection point and colorful region carry topological charge $Q=+1/2$ and $Q=+1$, respectively. In Fig.~\ref{fig:3}b, the topological charge
density of the $Q=3$ configuration at $(t_\perp=3, y_\perp=9)$ is plotted in the intersection plane, where we can identify the positive contributions indicated in Fig.~\ref{fig:2}b.

For the last step, we intersect two anti-parallel vortex pairs. The parameters for this vortex configuration are the same
as those of the $Q=1$ configuration and the angle profile of the vortex pairs is $\alpha_2$, given in Eq.~(\ref{eq:phi-pl0}). As shown in Fig.~\ref{fig:1}, for the sharp configuration, the total charge contribution $Q=0$ of four intersection points appears. The topological charge after smoothing the configuration approaches from $Q=0$ to $Q=0-1=-1$. The schematical diagram for the intersection plane of this configuration is shown in Fig.~\ref{fig:2}c. Two of the intersection points carry topological charge $Q=+1/2$ while the other two intersection points have $Q=-1/2$ and the colorful region carries topological charge $Q=-1$. They sum up to total topological charge $Q=-1$. In Fig.~\ref{fig:3}c, the topological charge density of the $Q=-1$ configuration at $(t_\perp=3, y_\perp=9)$ is plotted in the intersection plane, where we can identify the positive and negative contributions indicated in Fig.~\ref{fig:2}c.

In the next section, we analyze the low-lying modes of the overlap Dirac
operator in the background of these vortex configurations.

\section{Fermionic Dirac modes for the vortex fields}\label{Sect3}

In the previous section, we defined some vortex configurations which are combinations of the color structures and intersection points. For studying the effect of these configurations on fermions $\psi$, we determine the low-lying chiral eigenvectors $\chi_{R,L}$ and eigenvalues $|\lambda| \in [0,2/a]$ of the overlap Dirac operator \cite{Narayanan:1993ss,Narayanan:1994gw,Neuberger:1997fp,Edwards:1998yw}
\begin{eqnarray}\label{eq:ov_dirac}
D_{ov}=\frac{1}{a}\left[1+ \gamma_5 \frac{H}{|H|}\right]
\textrm{ with }H=\gamma_5 A,\;A=a D_\mathrm{W}-m,
\end{eqnarray}
where $D_W$ denotes the massless Wilson Dirac operator \cite{Wilson:1974sk,Gattringer:2010zz}. The mass parameter $m$ which is in the range $(0,2)$, is chosen with $m=1$ and the lattice constant $a$ is set to $a=1$. The eigenvalues of the overlap Dirac operator as a Ginsparg-Wilson operator are restricted to a circle in the complex plane.  This circle crosses the real axis at the two points $
\lambda= 0$ and $\lambda=1$ (doubler modes). Zero and doubler modes have exact chirality {\it i.e.} they are eigenvectors of $\gamma_5$. The eigenvalues $|\lambda| \in (0,2/a)$ come in complex conjugate pairs. Both of the eigenvectors $\psi_\pm=\frac{1}{\sqrt 2} (\chi_R \pm i\chi_L)$ belonging to one value of $|\lambda|\neq 0,2/a$
have the same chiral density as
\begin{equation}\label{eq:rho}
\chi=\psi^\dagger_\pm\gamma_5\psi_\pm=\frac{1}{2}(\chi_R^\dagger\chi_R-\chi_L^\dagger \chi_L).
\end{equation}

According to the Atiyah-Singer index theorem, the topological charge has to be related to zero modes of the Dirac operator by $\mathrm{ind}D[A]=n_--n_+=Q$ where $n_-$ and $n_+$ denote the numbers of left- and right-handed zero modes 
\cite{Atiyah:1971rm,Schwarz:1977az,Brown:1977bj,Adams:2000rn}. For a single configuration, one never finds fermionic zero modes of both chiralities. It means that for a gauge field with non-vanishing topological charge $Q$, the overlap Dirac
operator $D_{ov}$ has $|Q|$
exact zero modes with chirality $-\mathrm{sign} Q$. Therefore, any source of topological charge can attract zero modes contributing through interactions to a finite
density of near-zero modes leading to chiral symmetry breaking via the Banks-Casher relation. 

As mentioned above, we defined the $Q=1$ ($Q=3$) configuration which is two intersecting parallel vortex pairs where one of the vortices is colorful, negatively (positively) charged and also the $Q=-1$ configuration which is two intersecting anti-parallel vortex pairs where one of the vortices is colorful,  negatively charged.

\begin{figure}[h!]
\centering
\includegraphics[width=0.45\columnwidth]{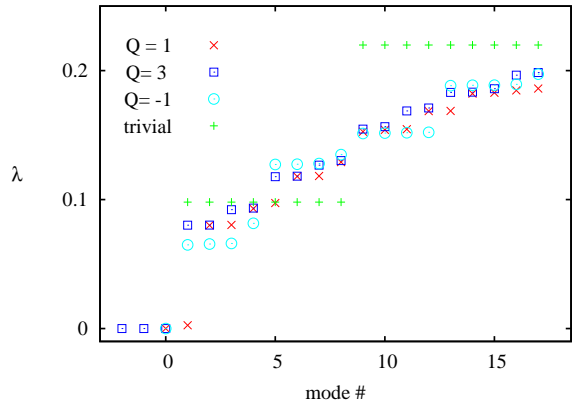}
\caption{ The lowest overlap eigenvalues for the vortex configurations schematically depicted in Fig.~\ref{fig:2} with $Q = 1$ (Fig.~\ref{fig:2}a), $Q = 3$ (Fig.~\ref{fig:2}b), and $Q = -1$ (Fig.~\ref{fig:2}c), compared with those of the free Dirac operator on a $16^4$ lattice. }
\label{fig:4}
\end{figure}

In Fig.~\ref{fig:4}, we show the lowest eigenvalues of the overlap Dirac operator in the background of these configurations compared to the eigenvalues of the free
overlap Dirac operator. For the fermion fields, we use anti-periodic boundary conditions in temporal direction and periodic boundary conditions in spatial directions on a $16^4$-lattice. The parameters of the configurations are the same as those in the previous section (given in Fig. ~\ref{fig:3}). For better comparison, we indicate the Dirac modes with mode number $\#(m)$ ($m\leqslant0$ means zero modes and $m>0$ would be non-zero modes). As shown in Fig. ~\ref{fig:4}, we find one zero mode $\#(0)$ of negative chirality for the $Q = 1$ configuration, three zero modes, $\#(-2)$, $\#(-1)$, $\#(0)$ of negative chirality for the $Q = 3$ configuration and one zero mode $\#(0)$ of positive chirality for the $Q = -1$ configuration, as expected from the index theorem. As shown in Fig.~\ref{fig:4}, we also get one near-zero mode $\#(1)$ for the $Q = 1$ configuration. The near-zero mode $\#(1)$ is distinguished from the mode $\#(0)$ by the chirality properties,
making it a near-zero mode as opposed to a zero mode.
 
\begin{figure}[h!]
\centering
a)\includegraphics[width=0.4\columnwidth]{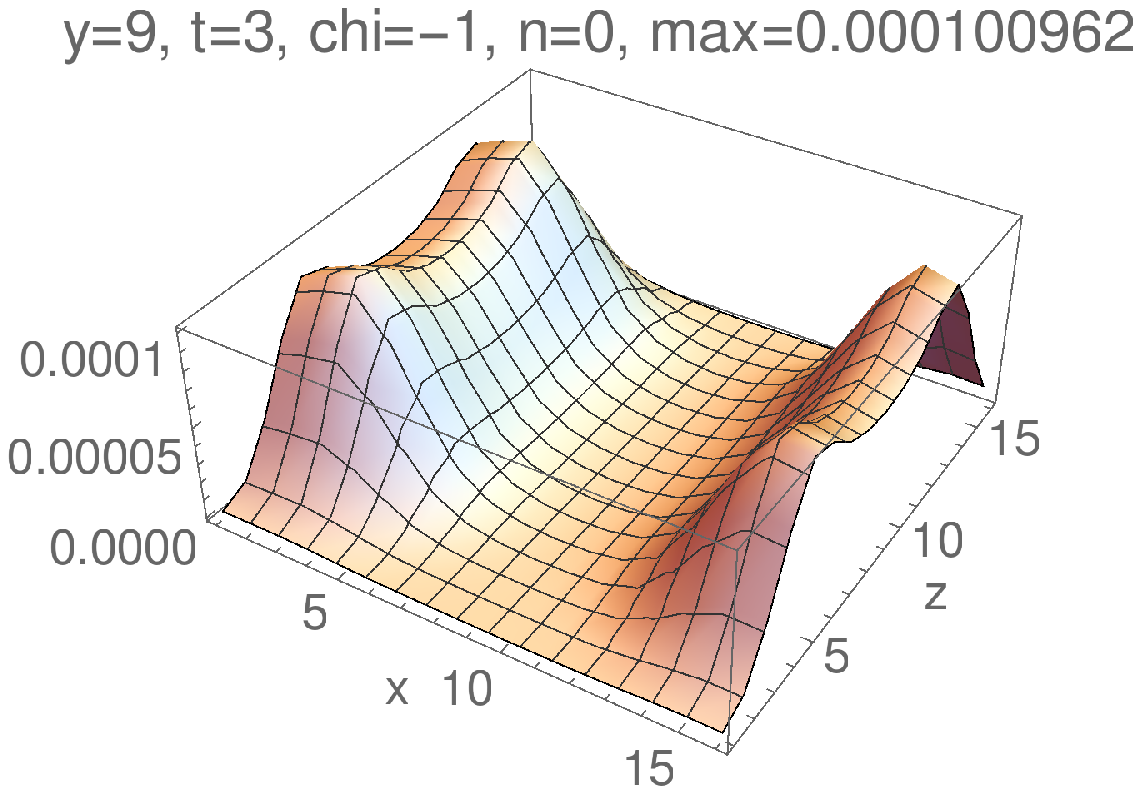}
b)\includegraphics[width=0.4\columnwidth]{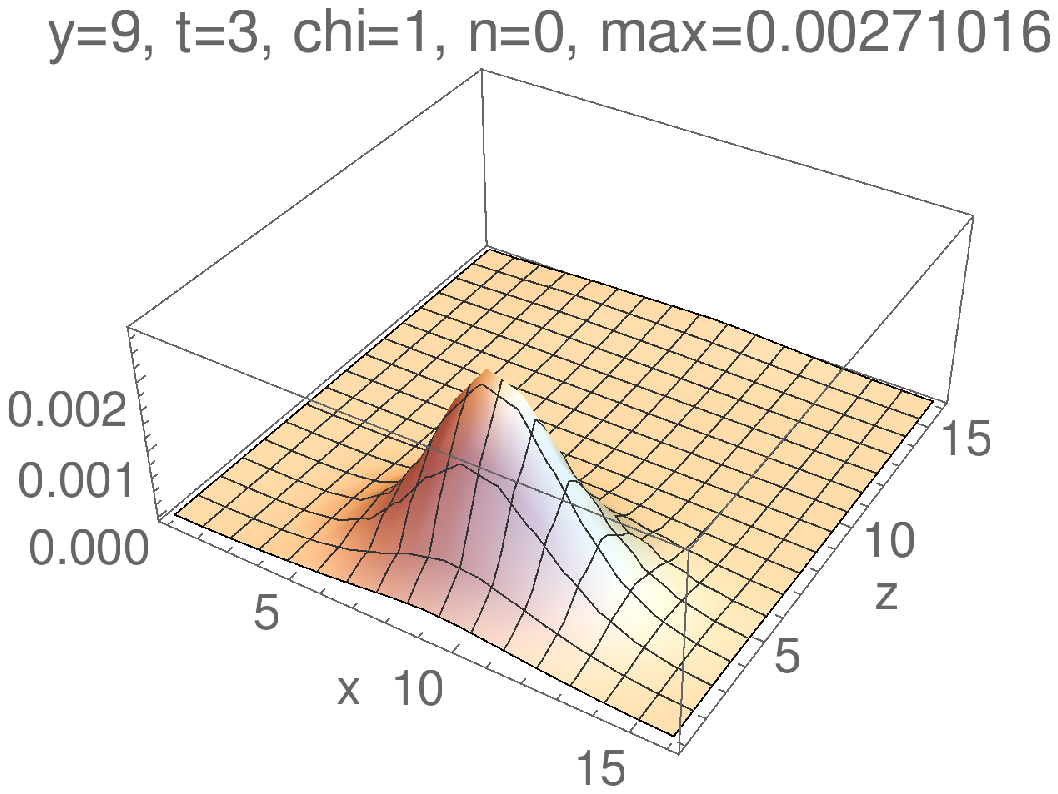}
\caption{ The chiral density $\chi(x)$ of the zero mode a) for the $Q = 1$ configuration b) for the $Q = -1$ configuration in the xz-plane. For the $Q = 1$ configuration, the chiral density of the zero mode shows four distinct maxima near to the intersection points. The colorful region in the configuration do not attract the chiral density of the zero mode any more. For the $Q = -1$ configuration, the chiral density of the zero mode is localized in the colorful region.}
\label{fig:5}
\end{figure}

\begin{figure}[h!]
\centering
a)\includegraphics[width=0.31\columnwidth]{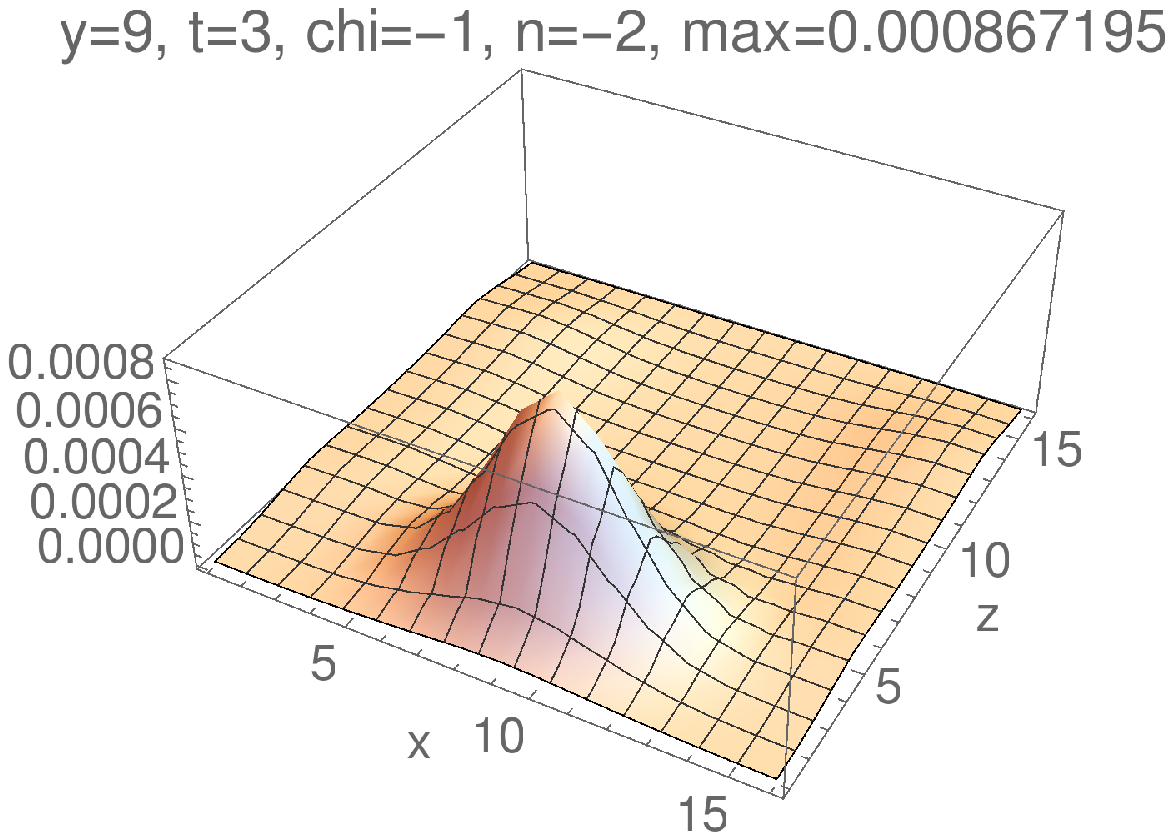}
b)\includegraphics[width=0.31\columnwidth]{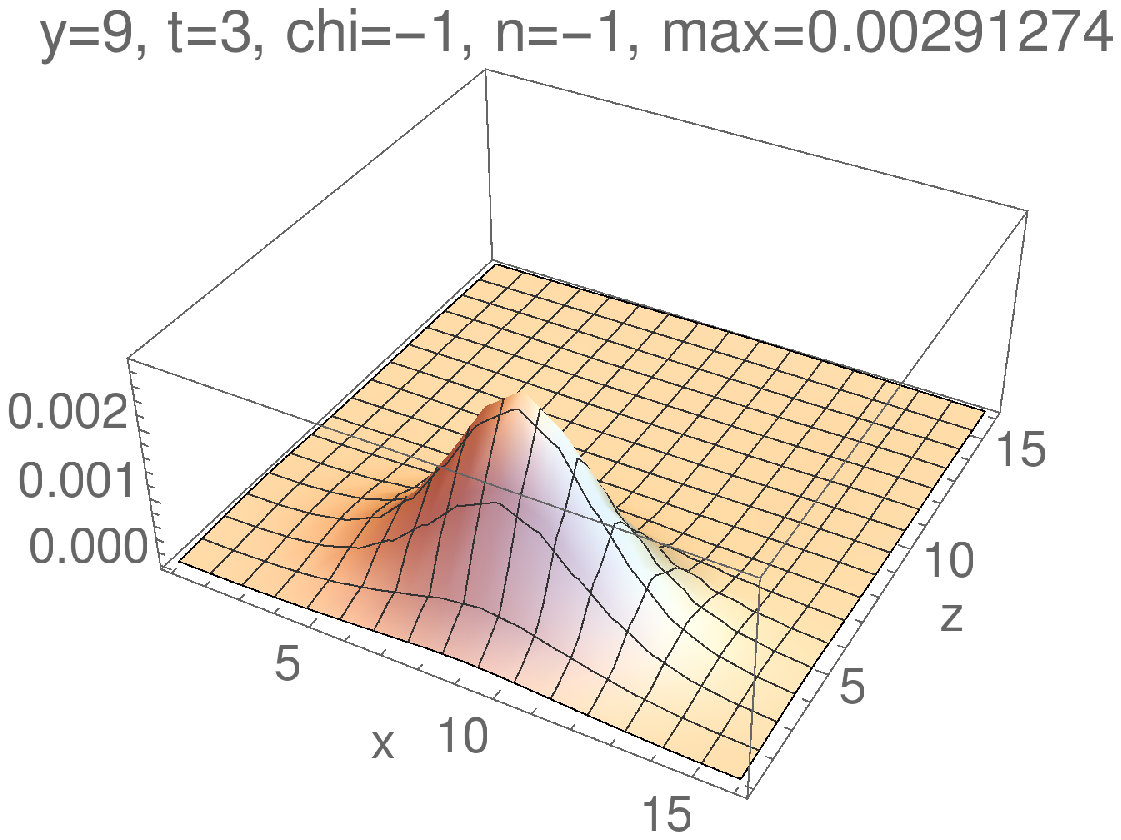}
c)\includegraphics[width=0.31\columnwidth]{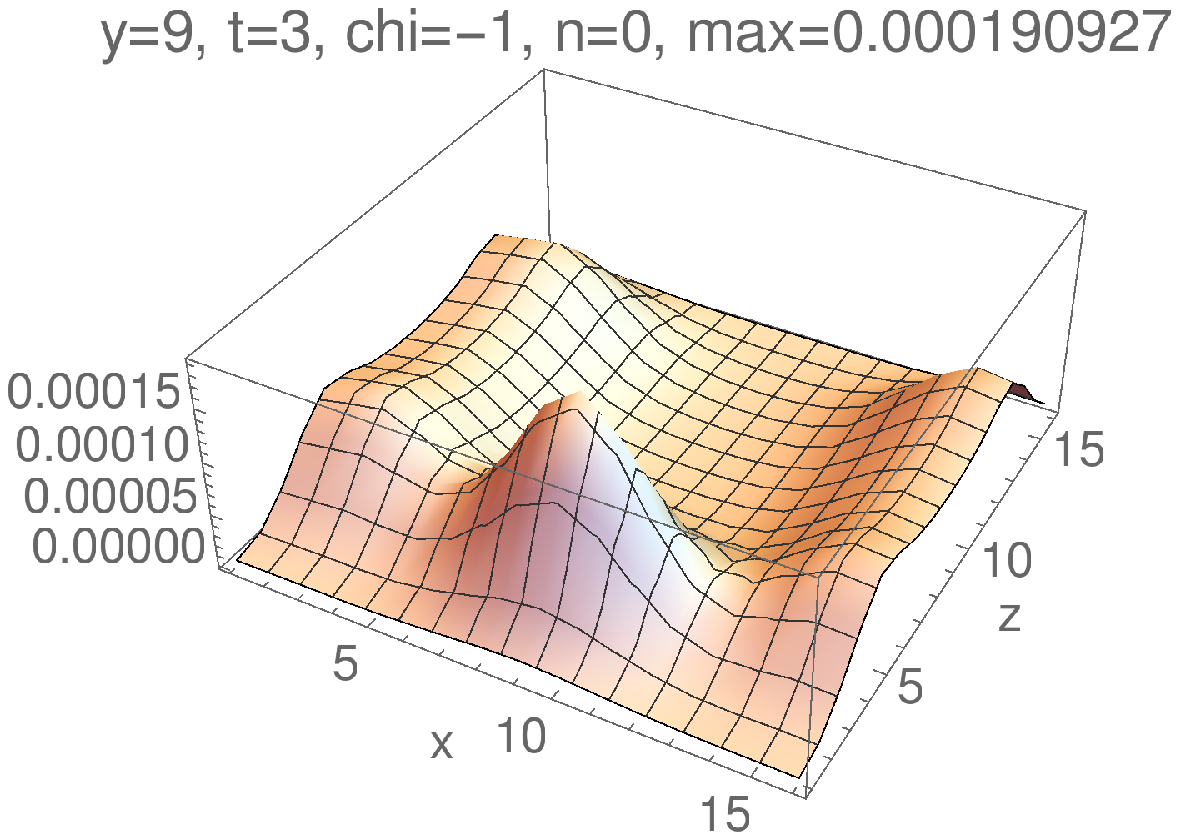}
\caption{ The chiral densities of three zero modes for the $Q = 3$ configuration. The chiral densities of the zero modes peak at the center of the colorful region. Two zero modes $\#(-2)$ and $\#(-1)$ are concentrated completely on the colorful region and one zero mode $\#0$ on both intersection and colorful regions. }
\label{fig:6}
\end{figure}

Now, we study the properties of zero modes and near-zero modes. In Fig.~\ref{fig:5} and Fig.~\ref{fig:6} we show the chiral densities of the zero modes for these configurations. The plot titles in the density plots indicate the plane positions, the chirality (``chi=0'' means we plot $\chi$, given in Eq.~(\ref{eq:rho}), ``chi=1'' would be right-handed chiral density $\chi=\frac{1}{2}(\chi_R^\dagger \chi_R)$ and ``chi=-1'' is left-handed chiral density $\chi=\frac{1}{2}(\chi_L^\dagger \chi_L)$), the number $n$ is related to the plotted mode $\#(n)$ and the maximal density in the plotted area, ``max=...''. 

For the $Q = 1$ configuration, the total topological charge is a combination of $Q = 2$ (intersection contributions) and $Q =-1$ (color structure contribution). As shown in Fig.~\ref{fig:5}a, the chiral density $\chi(\#0)$ of the zero mode for the $Q = 1$ configuration shows four distinct maxima near to the intersection points. Although there is the colorful region in the configuration, but it do not attract the chiral density $\chi(\#0)$ of the zero mode any more. 

For the $Q =-1$ configuration, the total topological charge is a combination of $Q=0$ (intersection contributions) and $Q=-1$ (color structure contribution). As shown in Fig.~\ref{fig:5}b, the chiral density $\chi(\#0)$ of the zero mode for the $Q = -1$ configuration is localized in the colorful region. 

For the $Q = 3$ configuration, the total topological charge is a combination of $Q = 2$ (intersection contributions) and $Q = 1$ (color structure contribution). In Fig.~\ref{fig:6}, we show the chiral densities of three zero modes for the $Q = 3$ configuration. Although a naive identification of topological and chiral densities would suggest that there should be one zero mode concentrated on
the colorful region and two on the intersection points, a quite different
distribution is in fact observed. As shown, the chiral densities of the zero modes peak at the center of the colorful
region. Two zero modes $\#(-2)$ and $\#(-1)$ are concentrated completely on the colorful region and one zero mode $\#0$ on both intersection and colorful regions. 

The index theorem may mislead one to identify topological densities with
chiral densities and deduce from the location of the chiral density where
topological density should be assigned. However, the index theorem is only a statement about integrated densities, not
the densities themselves. The densities cannot be identified one-to-one.

Considering Figs. \ref{fig:5} and \ref{fig:6} together, it seems that one can conclude that zero modes are correlated more strongly with the colorful vortex region than with the intersection points whenever the chirality of the zero mode can match the sign of either of those regions of topological charge. Only in Fig. \ref{fig:5}a, the case which is constructed such that the chirality of the zero mode only matches the sign of the topological charge due to intersection points, the zero mode remains attached to the intersection points.

Now, we study the low-lying modes of the considered vortex configurations after zero modes. As mentioned above, we find one near-zero mode $\#(1)$ for the $Q = 1$ configuration. For the $Q = 1$ configuration, the total topological charge is a combination of $Q = 2$ (intersection contributions) and $Q = -1$ (color structure contribution). The Dirac operator in the background of this configuration has one negative chirality zero mode (tied to a positive, intersection, contribution $Q=1$) and two would-be zero modes (for another intersection contribution $Q=1$ and the color structure contribution $Q=-1$). The near-zero mode of the $Q = 1$ configuration, as shown in Fig.~\ref{fig:4}, originates from the overlap of these two would-be zero modes. The left-handed chiral density $\chi(\#1)$ (chi=-1) of the near-zero mode behaves similar to the chiral density (chi=-1) of the zero mode $\chi(\#0)$ for the configuration $Q = +1$ (see Fig.~\ref{fig:5}a). The right-handed chiral density $\chi(\#1)$ (chi=1) of the near-zero mode behaves similar to the chiral density (chi=1) of the zero mode $\chi(\#0)$ for the $Q = -1$ configuration (see Fig.~\ref{fig:5}b). As a result, the left-handed chiral density of the near-zero mode spreads near to four intersection points while the right-handed chiral density of the near-zero mode is localized near to the colorful region. 

\begin{figure}[h!]
\centering
a)\includegraphics[width=0.22\columnwidth]{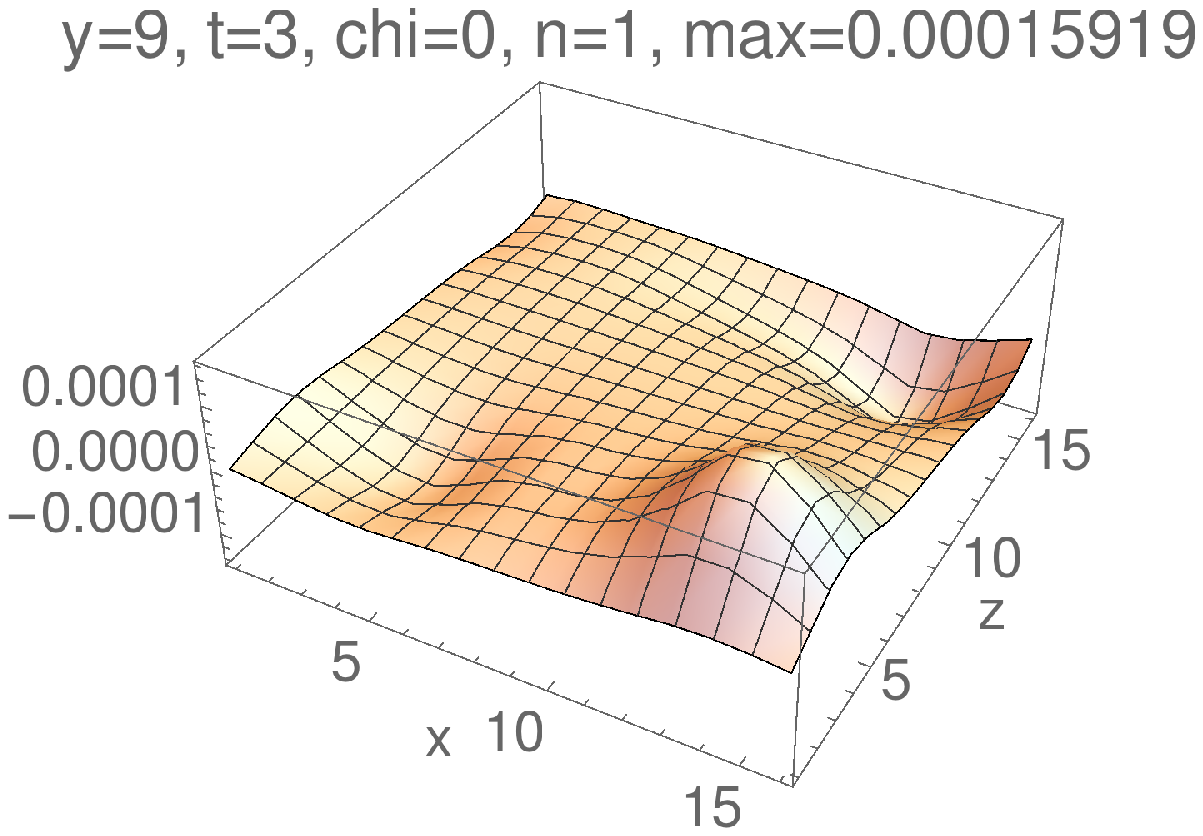}
b)\includegraphics[width=0.22\columnwidth]{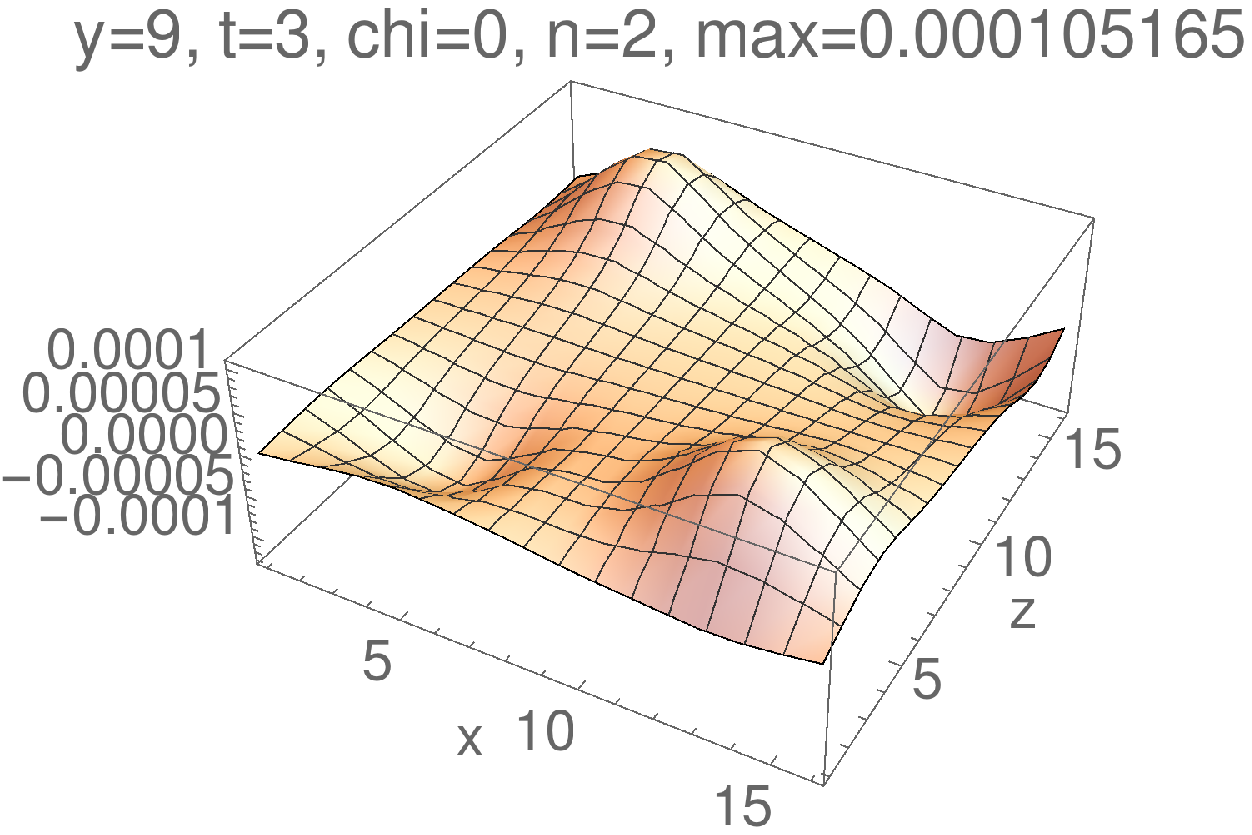}
c)\includegraphics[width=0.22\columnwidth]{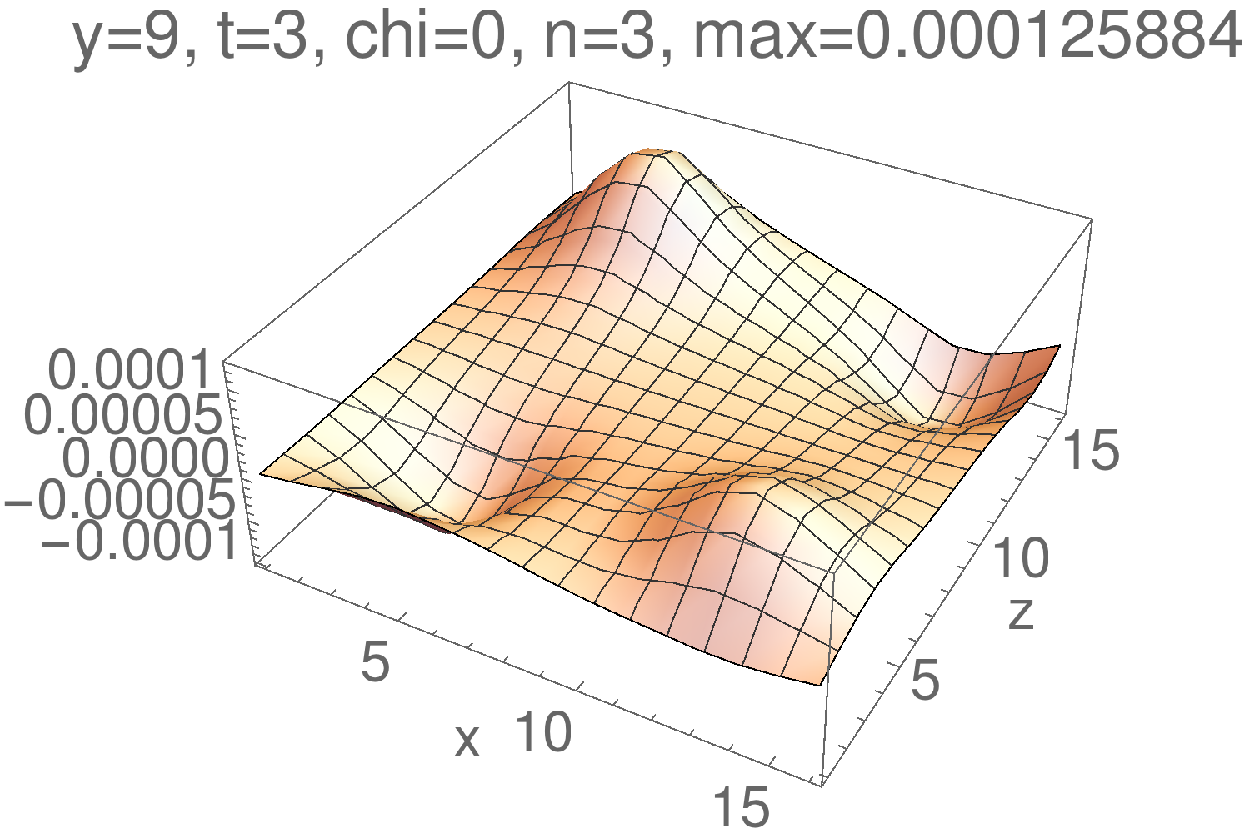}
d)\includegraphics[width=0.22\columnwidth]{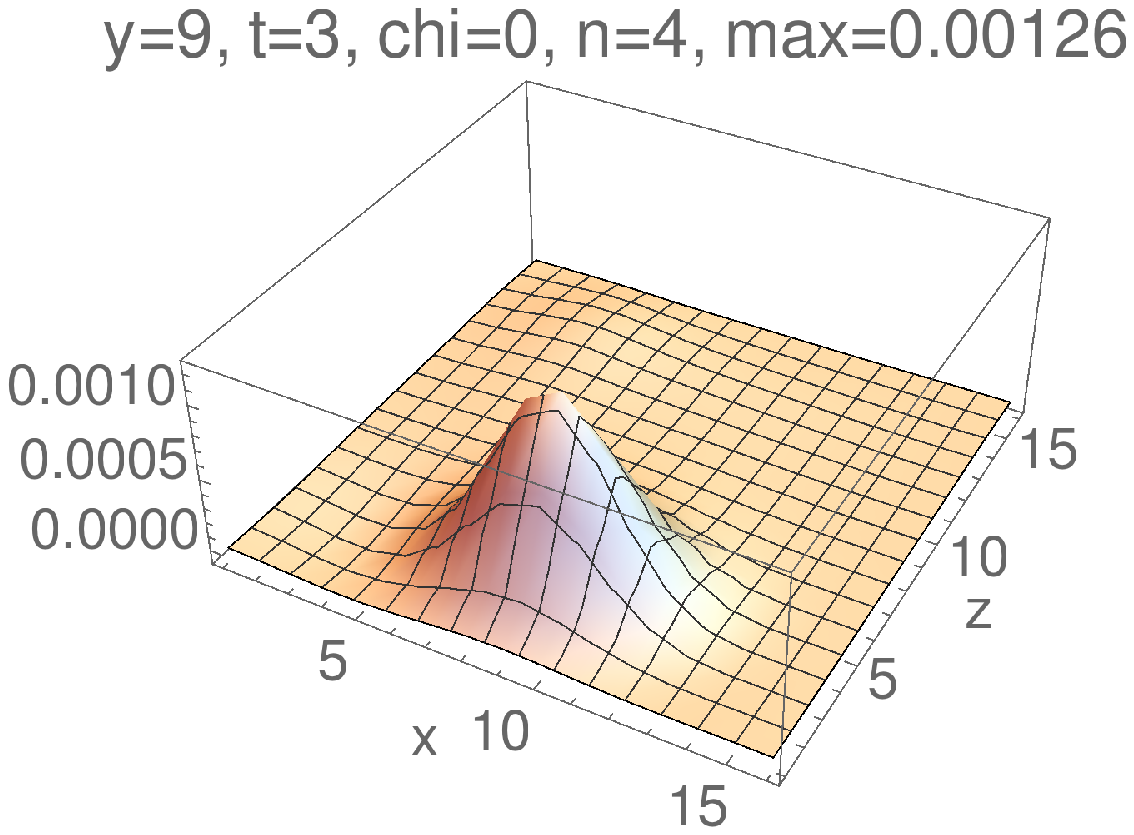}
\caption{ The chiral densities of four low-lying modes after one zero mode for the $Q = -1$ configuration. The first three chiral densities a), b), and c) are localized near to intersection points. d) The fourth mode is localized near to the colorful region. }
\label{fig:7}
\end{figure}
\begin{figure}[h!]
\centering
\includegraphics[width=0.45\columnwidth]{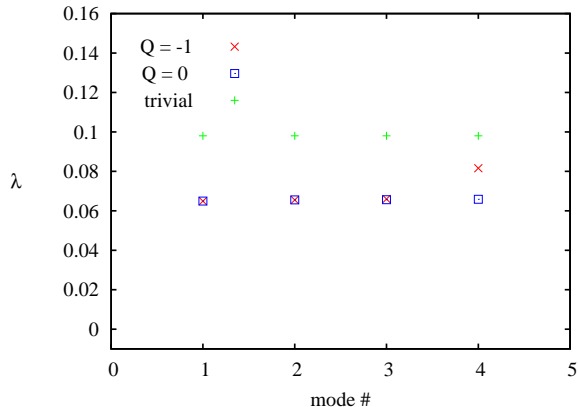}
\caption{ The four low-lying eigenvalues for the $Q = -1$ and $Q = 0$ configurations compared with those of the free Dirac operator. For the $Q = 0$ configuration, two anti-parallel vortex pairs are intersected. The $Q = -1$ configuration is the same as $Q = 0$ configuration but we have added to the configuration the colorful region with $Q = -1$. The eigenvalues of the first three eigenmodes for the $Q = 0$ configuration are the same as those of $Q = -1$ configuration. The eigenvalue of the fourth low-lying mode of the $Q = 0$ configuration is increased after locating within the colorful region.}
\label{fig:8}
\end{figure}

Now, for the $Q = -1$ configuration, we study four low-lying modes, $\#(1)$, $\#(2)$, $\#(3)$, $\#(4)$ as shown in Fig.~\ref{fig:4}. For the $Q = -1$ configuration, the total topological charge is a combination of $Q = 0$ (intersection contributions) and $Q = -1$ (color structure contribution). The $Q = 0$ configuration, as an intersecting unicolor center vortex field with $Q=0$ have been studied in Ref. \cite{Hollwieser:2011uj} where two of the intersection points carry topological charge $Q=+1/2$ while the other two intersection points have $Q=-1/2$. Two anti-parallel vortex pairs are intersected similar to the $Q = -1$ configuration but we do not have the colorful region. For the $Q = 0$ configuration, the first four low-lying modes are localized at intersection points. The chiral densities of four low-lying modes $\chi(\#1)$, $\chi(\#2)$, $\chi(\#3)$, $\chi(\#4)$ of the $Q = -1$ configuration are shown in Fig.~\ref{fig:7}. The chiral densities of the three low-lying modes $\chi(\#1)$, $\chi(\#2)$, $\chi(\#3)$ are localized at intersection points while $\chi(\#4)$ (5th mode) is localized in the colorful region. In Fig.~\ref{fig:8}, the eigenvalues of these four low-lying modes of the $Q = -1$ configuration are compared with those of the $Q = 0$ configuration. The low-lying modes $\#(1)$, $\#(2)$, $\#(3)$ for both configurations have the same eigenvalues while the eigenvalue of the low-lying mode $\#(4)$ is increased compared with the one of the fourth low-lying mode of the $Q = 0$ configuration. Therefore, the eigenvalue of the mode $\#(4)$ of the analogous $Q=0$ configuration is increased through the influence of the colorful region. It is possible that there is a level crossing and the mode $\#(4)$ of the $Q=0$ configuration, after distortion by the colorful region, becomes a higher
mode of the $Q=-1$ configuration.

\section{Conclusion}\label{Sect4}

Understanding the dynamical mechanism of the spontaneous breaking of chiral symmetry in QCD is our aim. We analyze special vortex configurations which are combinations of intersection points and colorful regions for studying the properties of low-lying modes of the Dirac operator. We construct vortex configurations by combining two perpendicular plane vortex pairs and making one of the plane vortices colorful. The intersection points and the colorful region are apart. Our special vortex configurations have total topological charge $Q=\pm1$ and $Q=3$. $Q=1$ ($Q=3$) configuration is two intersecting parallel vortex pairs where one of the vortices is colorful, negatively (positively) charged and also $Q=-1$ configuration is two intersecting anti-parallel vortex pairs where one of the vortices is colorful, negatively charged. 

These configurations give the nice opportunity to study the properties of zero modes and near-zero modes of the overlap Dirac operator. We have analyzed the low-lying modes of the overlap Dirac operator in the background of these configurations. The data provide insight, in particular, into the interplay between chiral densities induced by vortex intersection points and by color structure. 

For the $Q = 1$ configuration, we find one zero mode of negative chirality and one near-zero mode. The chiral density of the zero mode shows four distinct maxima near to the intersection points. The chirality of the zero mode only matches the sign of the topological
charge due to intersection points, the zero mode remains attached to the intersection points. The Dirac operator in the background of this configuration has also two would-be zero modes (for another intersection contribution $Q = 1$ and the color structure contribution $Q = -1$). The near-zero mode originates from the overlap of these two would-be zero modes. The left-handed chiral density of the near-zero mode spreads near to four intersection points while the right-handed chiral density of the near-zero mode is localized near to the colorful region.

For the $Q = -1$ configuration, we find one zero mode of positive chirality. The chiral density of the zero mode is localized in the colorful region. For this configuration, the total topological charge is a combination of $Q = 0$ (intersection contributions) and $Q = -1$ (color structure contribution). For the analogous unicolor $Q=0$ configuration,
the chiral densities of the first four low-lying modes are localized in the intersection points. For the $Q=-1$ configuration, the chiral densities of three of the first four low-lying modes, after one zero mode, behave similarly, but the chiral density of the fourth mode is localized in the colorful region. The eigenvalue of this mode is increased compared to the fourth low-lying mode of the $Q=0$ configuration.

For the $Q = 3$ configuration, we find three zero modes of negative chirality. Although a naive identification of topological and chiral
densities would suggest that there should be one zero mode concentrated on
the colorful region and two on the intersection points, a quite different
distribution is in fact observed. The chiral densities of three zero modes peak at the center of the colorful region. Two zero modes are concentrated completely on the colorful region and one zero mode on both intersection and colorful regions.

It seems that one can conclude that zero modes are correlated more strongly with the colorful vortex region than with the intersection points whenever the chirality of the zero mode can match the sign of either of those regions of topological charge. 

Zero modes and near-zero modes of the type studied in this work may be instrumental in generating a non-zero spectral density of the Dirac operator nearzero eigenvalue, and thus lead to spontaneous chiral symmetry breaking.

\section{\boldmath Acknowledgments}
 I am grateful to Manfried Faber, Roman H\"ollwieser, and Urs M. Heller to prepare the basic programs. 

\bibliographystyle{unsrt}
\bibliography{chiral}

\end{document}